%% file: ms.tex
\documentclass{emulateapj}
\usepackage{natbib}
\usepackage{amsmath,  amsthm, amssymb, cases}
\newcommand{\ts}{t_\mathrm{stop}}

\newcommand{\gcc}{\;\mathrm{g/cm^{3}}}

\newcommand{\vdr}{v_\mathrm{dr}}
\newcommand{\tdr}{t_\mathrm{dr}}
\newcommand{\tgr}{t_\mathrm{grow}}

\newcommand{\Sc}{\rm Sc}
\newcommand{\ag}{\alpha_g}
\newcommand{\amax}{\alpha_\mathrm{max}}
\newcommand{\kfgm}{k_\mathrm{f}}

\newcommand{\lfgm}{\lambda_\mathrm{f}}

\newcommand{\lmfp}{\lambda_\mathrm{mfp}}
\newcommand{\amm}{a_\mathrm{mm}}
\newcommand{\Qm}{Q_\mathrm{sonic}}
\newcommand{\torb}{t_\mathrm{orb}}

\slugcomment{Draft Modified \today}

\shorttitle{GI of Solids in Turbulent Disks}
\shortauthors{A.N. Youdin}

\begin{document}

\title{Planetesimal Formation without Thresholds. II: Gravitational Instability of Solids in Turbulent Protoplanetary Disks}
\author{Andrew N. Youdin}
\affil{Princeton University Observatory, Princeton, NJ 08544}

\begin{abstract}
We show that small solids in low mass, turbulent protoplanetary disks collect into self-gravitating rings.   Growth is faster than disk lifetimes and radial drift times for moderately strong turbulence, characterized by dimensionless diffusivities, $\alpha_g \lesssim 10^{-6}$---$10^{-3}$ when particles are mm-sized.  This range reflects a strong dependance on disk models.  Growth is faster for higher particle surface densities. Lower gas densities and larger solids also give faster growth, as long as aerodynamic coupling is tight.  In simple power law models, growth is slowest around $\sim 0.3$ AU, where drag coupling is strongest for mm-sized solids.  Growth is much faster close to the star where orbital times are short, with implications for \emph{in situ} formation of short period extrasolar planets.   Growth times also decrease toward the outer disk where lower gas densities allow greater particle settling.  Beyond roughly Kuiper Belt distances however, solids are sufficiently decoupled from gas that dissipative gravitational instabilities are less effective.   Turbulence not only slows growth, but also increases radial wavelengths.  The initial solid mass in an unstable ring can be $\sim 10^{-2} M_\oplus$ or greater, huge compared to kilometer sized planetesimals.  Nonlinear fragmentation, which has not been studied in detail, will lower the final planetesimal mass.  We consider applications to the asteroid belt and discuss the alternate hypothesis of collisional agglomeration.
\end{abstract}

\section{Introduction}
\subsection{Summary of Paper I}
\citet[hereafter Paper I]{y05a} analyzed the linear, axisymmetric gravitational instability of solids (GIS, or GI for gravitational instabilities in general) subject to gas drag.  Growth was described in terms of $\tau_s \equiv \Omega \ts$; the dimensionless measure of particle stopping time, $\ts$, where $\Omega$ is the orbital frequency; and two stability parameters:
\begin{eqnarray}
Q_T &\equiv& {c \Omega \over \pi G \Sigma}\, \label{QT},\\
Q_R &\equiv& {h \Omega^2 \over \pi G \Sigma} \approx {\Omega^2 \over \pi G \rho}\, \label{QR}.
\end{eqnarray}
The Toomre parameter, $Q_T$, measures the influence of particle velocity dispersions, $c$, in a Keplerian disk with particle surface density $\Sigma$, where $G$ is the gravitational constant.  The parameter $Q_R$ uses the particle sublayer thickness, $h$, to measure the ratio of the Roche density, $\sim \Omega^2/(\pi G)$, to the particle space density, $\rho = \Sigma/h$.  Paper I found that dissipation of angular momentum allows GIS for arbitrary values of the stability parameters, but growth is slower when they are large.

Paper I also derived values of $Q_T$ and $Q_R$ generated by turbulent stirring of solids.  Gas turbulence was characterized by a diffusive viscosity,
\begin{equation}\label{eq:nu}
\nu_g \equiv \ag c_g^2/\Omega \, ,
\end{equation}
where $c_g$ is the gas sound speed.  The dimensionless diffusivity, $\alpha_g$, is defined by analogy with the accretion disk literature, but it refers to diffusion of material, not angular momentum (see \S\ref{alphacomp} for a detailed discussion).  The analysis requires an assumption about the characteristic eddy turnover time, $t_0$.  Unless otherwise noted, we take $t_0 = 1/\Omega$, i.e. orbital turnover times.

The main equations from Paper I used in this work are briefly summarized.  Our general dispersion relation (eq. [13] of Paper I, hereafter eq. [I.13]) defines the dimensionless growth rate $\gamma$ (eq. [I.11]).  Wavenumbers, $\kfgm$, (or wavelengths, $\lfgm$) of the fastest growing modes satisfy equation (I.17).  Particle scale heights are determined by balancing vertical settling and turbulent diffusion (eq. [I.29]).  The velocity dispersion (eq. [I.30]) includes direct kicks from turbulent fluctuations (eq. [I.25]]) and a contribution from epicyclic motion (eq. [I.32]).  The resulting values of the stability parameters are plotted in Figure I.3 (i.e.\ Fig. 3 of Paper I), and limiting analytic expressions for $Q_R$ and $Q_T$ are given in equation (I.36). 
 
\subsection{This Work}
Since growth rates are always positive, our goal is to determine the conditions that give sufficiently fast growth times:
 \begin{equation} \label{eq:tgr}
 \tgr \equiv {1 / (\gamma \Omega)}\, .
 \end{equation}
Growth should be faster than the 1-10 Myr lifetimes of gas rich T Tauri disks \citep{acp03}, particularly if solid cores are to accrete massive gas atmospheres.  Since isotopic dating gives a 1-3 Myr spread in age of meteoritic inclusions \citep{ame02}, there is no indication that planetesimals must form much faster than this.  Second, growth should be faster than aerodynamic radial drift timescales \citep{stu77,nsh86}.  Also, the wavelength of unstable modes must be smaller than the disk radius, $R$.  We implement these conditions as:
\begin{subequations}\label{conditions}
\begin{eqnarray}
\tgr &<& 0.1 t_\mathrm{disk} \equiv 10^5 ~{\rm yrs} \label{diskcond} \, ,\\
\tgr &<& \tdr \equiv R/\vdr \label{driftcond} \, ,\\
\lfgm  &<& R/2 \label{lengthcond}\, ,
\end{eqnarray}
\end{subequations}
where the timescale for inward drift of solids, $\tdr$, is given in equation (\ref{tdr}).  These conditions may not be the only ones that allow planetesimal formation, but they describe the limits of validity of our local analysis of GIS.

This paper is organized as follows.  We present general, model independent results in \S\ref{mind}.  We define simple powerlaw disk models in \S\ref{sec:diskmodels}.  Growth for fixed $\alpha_g$ values is investigated in \S\ref{sec:fixalpha}.  The strongest levels of turbulence which allow GIS according to the above criteria (eqs. \ref{conditions}) are explored in \S\ref{sec:alpha}.   Comparison to expected levels of turbulence is made in \S\ref{sec:alphameaning}.  The results are discussed in \S\ref{sec:concl}.  Appendix \ref{sec:fasteddy} investigates the effect of fast eddy turnover times.

\input{tab1.tex}

\input{tab2.tex}

\begin{figure}[tb]
   \hspace{-.3in}  
\includegraphics[width=3.8in]{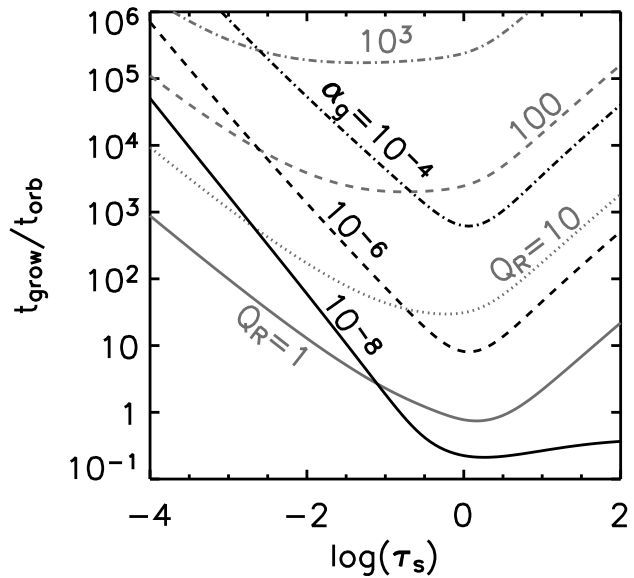} 
\caption{Growth time for gravitational collapse of solids vs.\ stopping time.  \emph{Black curves:} Turbulent diffusivity, $\ag$, held fixed.   \emph{Grey  curves:} $Q_R$, inversely proportional to particle space density, fixed.  Both curves are for orbital turnover times and $\Qm=5\times 10^3$.  See \S\ref{mindgrow}.}
   \label{fig:groworb}
   \vspace{.1cm}
\end{figure} 

\section{Model Independent Results}\label{mind}
We first describe the instability in terms of $\ag$ and $\tau_s$ without assuming a specific disk model.   We must specify a third parameter, $\Qm$ (eq. [\ref{QsonicN}]).\footnote{A two parameter description is possible with orbital turnover times since $\sqrt{\ag} \Qm$ always appears together.  We opt for the familiarity of ``$\alpha$ parameters" over a more compact description.}  We assume that eddies have orbital turnover times, but faster eddies are considered in Appendix \ref{sec:fasteddy}.  This analysis provides a quick understanding of the general behavior of unstable modes, but does not allow evaluation of the criteria in equations (\ref{conditions}).    

\subsection{Growth Times}\label{mindgrow}
Figure \ref{fig:groworb} plots growth times relative to the local orbital time, $\tgr/\torb = 2\pi/\gamma$.  The black curves hold  $\ag$ fixed.  The stability parameters $Q_R$ and $Q_T$ for this case were plotted in Figure I.3 (\emph{top}).  Growth is fastest for weak turbulent diffusion (small $\ag$) and marginal coupling, $\tau_s \approx 1$.  Tight coupling retards  collapse by slowing the terminal velocity of solids.  Loose coupling lowers the dissipation rate of angular momentum needed for collapse of long wavelength modes.  We emphasize that growth times are always finite.

Growth is also faster for higher particle surface densities, as self-gravity is stronger.  Increasing $\Sigma$ decreases growth times with a dependence between $\tgr \propto 1/\Sigma^2$ and $1/\Sigma$ for waves long and short (respecitively) compared to $h$.   This follows from the limiting analytic expressions in Table \ref{tab:limits} since $\Qm \propto 1/\Sigma$.  These expressions, derived from the results of Paper I (using eqs. [I.18, I.19, I.21, I.B2a]), explain the general features of the constant $\alpha_g$ curves.  The one exception is the rapid dynamical collapse for $\ag = 10^{-8}$ and $\tau_s > 1$.  In this case $Q_R \sim Q_T \lesssim 1$ and traditional GI dominates dissipative growth (see eq. [I.B1] and surrounding discussion).

Grey curves (in Fig.\ \ref{fig:groworb}) plot growth at fixed $Q_R$.  Growth in under $10^5$ orbits is possible for $Q_R \lesssim 800$, densities nearly a thousand times smaller than the Roche density.   Compared to the standard $Q_R \lesssim 1$ criteria, dissipation allows GIS for much stronger turbulence.

\begin{figure}[tb!] 
    \hspace{-.3in} 
 \includegraphics[width=3.8in]{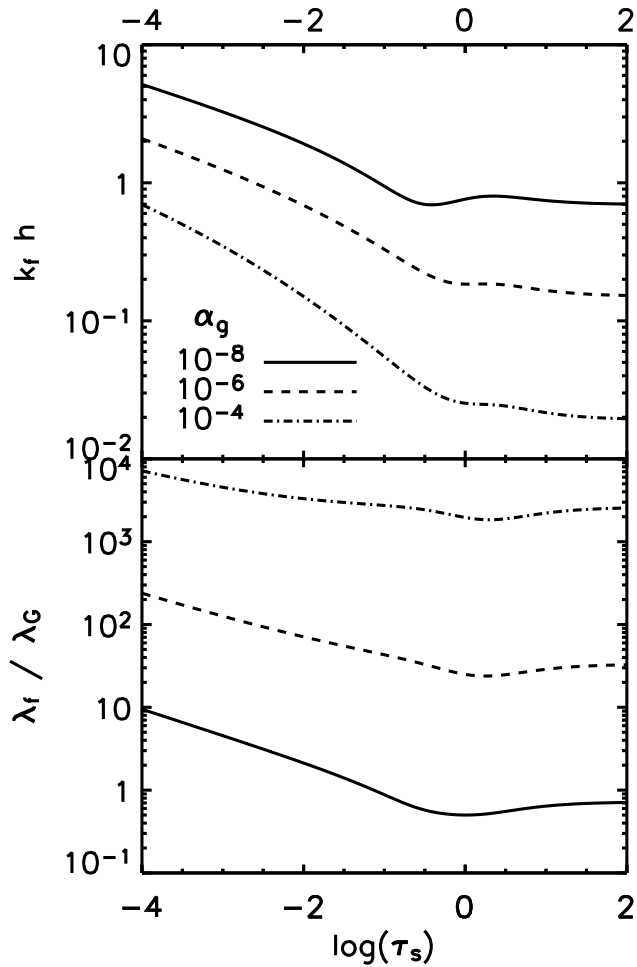} 
   \caption{\emph{Top:} fastest growing wavenumber times scale height vs.\ stopping time for fixed $\ag$.  \emph{Bottom:} most unstable wavelength compared to the standard GI wavelength.   Both plots assume orbital turnover times and use the same legend for $\ag$ values.}
   \label{fig:kl}
\end{figure}

\subsection{Radial Wavelengths}
Figure \ref{fig:kl} (\emph{bottom}) compares the most unstable wavelength, $\lfgm  = 2 \pi/\kfgm$ to the standard (non-dissipative) wavelength of GI, 
\begin{equation}\label{lg}
\lambda_G \equiv 2 \pi^2 G \Sigma/\Omega^2.
\end{equation}
We find $\lfgm  \gg \lambda_G$ unless turbulence is very weak or (consulting Tab. \ref{tab:limits}) $\Sigma$ is very large.  Such long wavelengths are allowed by angular momentum dissipation.  The initial ring mass $\sim 2\pi \Sigma \lfgm R$ vastly exceeds the traditional estimate $\sim \Sigma \lfgm^2$ corresponding to km-sized planetesimals  \citep{gw73}, but subsequent fragmentation is likely (see \S\ref{sec:slowmodes}).  The weak dependence on $\tau_s$ bodes well for studies that will include a dispersion of particle sizes.  A detailed analysis is needed, but a range of particle sizes could collect in the same annulus with larger solids (i.e.\ closer to $\tau_s = 1$) collapsing faster.

Figure \ref{fig:kl} (\emph{top}) plots $\kfgm h$, measuring the size of waves relative to the layer thickness.\footnote{Since $\lfgm /\lambda_G = Q_R/(\kfgm h)$, the two plots in Figure \ref{fig:kl} differ only by a factor of $Q_R$.}    Since $\kfgm h < 2 \pi$, our vertically integrated model is a good approximation.   Our limiting analytic formulae (Tab. \ref{tab:limits}) apply for $\kfgm h \ll 1$ or $\gg 1$.  Figure \ref{fig:kl} indicates which case is more applicable.  When $\kfgm h \sim 1$, behavior is intermediate and requires numerical evaluation.

\begin{figure}[htb]
   \hspace{-.3in} 
\includegraphics[width=3.8in]{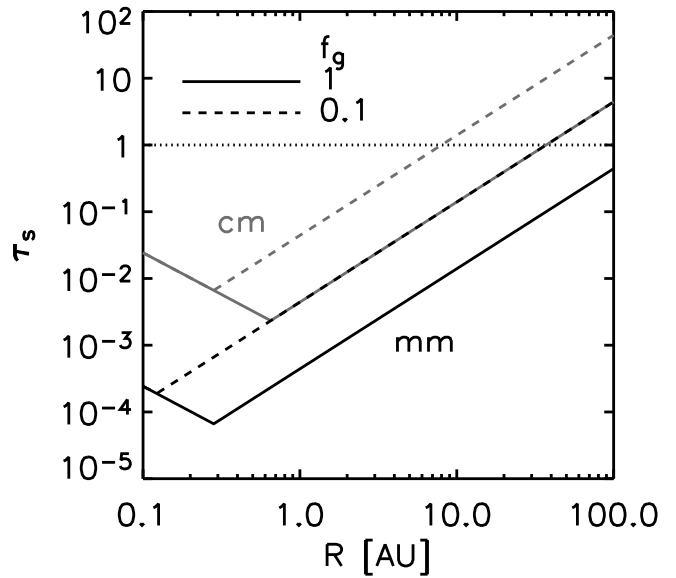} 
   \caption{Dimensionless stopping time vs.\ disk radius for  mm (black) and  cm (grey) sized solids with a material density of 3 gm/cm$^3$ in a standard gas disk (solid lines) and with gas depleted (dashed  lines).  Marginal coupling is indicated (dotted line).  Kinks demark the transition between Stokes (small $R$) and Epstein (larger $R$) drag, which in reality would be smoothed.}
   \vspace{.2cm}
   \label{fig:ts}
\end{figure} 

\section{Protoplanetary Disk Model}\label{sec:diskmodels}
Our numerical analysis  uses a standard minimum mass model (\citealp{stu77b}; \citealp{hay81}; our notation   follows \citealp{ys02}, hereafter YS02) for the particle and gas surface densities and the sound speed: 
\begin{eqnarray}
\Sigma &=& f_p 10 \varpi^{-3/2} \gcc \\
\Sigma_g &=&  f_g 1700 \varpi^{-3/2} \gcc \\
c_g &=& 1 \varpi^{-1/4} ~{\rm km/s}
\end{eqnarray}
where $\varpi \equiv R/{\rm AU}$ gives stellocentric distance in AU, and $f_p$ and $f_g$ are enhancement (or depletion) factors for solids and gas, respectively.  Our reference model has $f_g = f_p = 1$.  We treat $f_p$ and $f_g$ as constants to understand the behavior of simple power law models.   However many enrichment or depletion mechanisms deviate signficantly from power law behavior, including:
ice condensation (many models give $f_p = 4$ outside the snow-line),  particle pile-ups generated by inward radial drift of solids (YS02; \citealp{yc04}, hereafter YC04), and photoevaporation of gas \citep{tb05}.  

Particles are assigned a uniform size $a$ (1 mm unless stated otherwise) and density $\rho_s = 3 \gcc$.  These are typical properties of the abundant meteoritic inclusions called chondrules.  It is a realistic concern that a dispersion of particle sizes might yield less efficient collapse.  We plan to study this issue in the future, but note that chondrules are strongly size-sorted \citep[p. 117]{tay01}.

For $a \lesssim (9/4) \lmfp$, where \begin{equation}\label{mfp}
\lmfp = 1 f_g^{-1} \varpi^{2.75}~{\rm cm},
\end{equation}
is the gas mean free path, the stopping time is given by Epstein's law: 
\begin{equation}\label{Ep}
\tau_s^{\rm Ep} = \sqrt{2\pi} {\rho_s a/\Sigma_g} \approx 4 \times 10^{-4} \amm {\varpi^{3/2} / f_g},
\end{equation}
where $\amm = a/(1~{\rm mm})$.  Unless stated otherwise, numerical estimates will use Epstein's law.  For $a > (9/4) \lmfp$, i.e.\ small Knudsen numbers, Stokes' drag law applies and
\begin{equation}\label{St}
\tau_s^{\rm St} = \sqrt{2\pi} {4 \rho_s a^2 \over 9 \Sigma_g \lmfp} \approx 9 \times 10^{-5}\amm^2\left({\varpi\over 0.3}\right)^{-5/4},
\end{equation}
independent of gas density.  Figure \ref{fig:ts} plots the stopping time for both mm and cm-sizes.  The effect of gas depletion (to $f_g = 0.1$) is demonstrated.  Since $\tau_s^{\rm St}$ decreases with disk radius, while $\tau_s^{\rm Ep}$ increases, a given particle is most tightly coupled at the transition between the two regimes.  We do not consider particles large enough for turbulent drag laws to be relevant.    

Other relevant quantities take the following values
\begin{eqnarray}
\Qm &\equiv& c_g \Omega/(\pi G \Sigma) \approx 9.5 \times 10^3 \varpi^{-1/4}/f_p\, \label{QsonicN},\\
Q_g &\equiv& c_g\Omega/(\pi G \Sigma_g) \approx 56 \varpi^{-1/4}/f_g\, ,\\
\eta &\equiv& -{\partial P / \partial R \over 2 \Omega^2 R \rho_g} \approx 1.8 \times 10^{-3} \sqrt{\varpi}\, \label{eta},\\
\tdr &=& {1+\mu_g^2\tau_s^2 \over 2 \mu_g^2\eta \Omega \tau_s} \approx {10^5 f_g \over \mu_g^2 \amm \sqrt{\varpi}} ~\mathrm{yrs} \label{tdr},
\end{eqnarray}
where $\mu_g = \rho_g/(\rho_g + \rho)$ is the midplane gas fraction.  Large $Q_g$ is consistent with our assumption that the gas disk as a whole is not gravitationally unstable.  The numerical expression for $\tdr$ assumes tight coupling and Epstein drag.  Drift times are longer than nominal values if particle inertia dominates in the midplane to give $\mu_g < 1$.

\begin{figure}[tbp]
   \hspace{-.3in} 
 \includegraphics[width=3.8in]{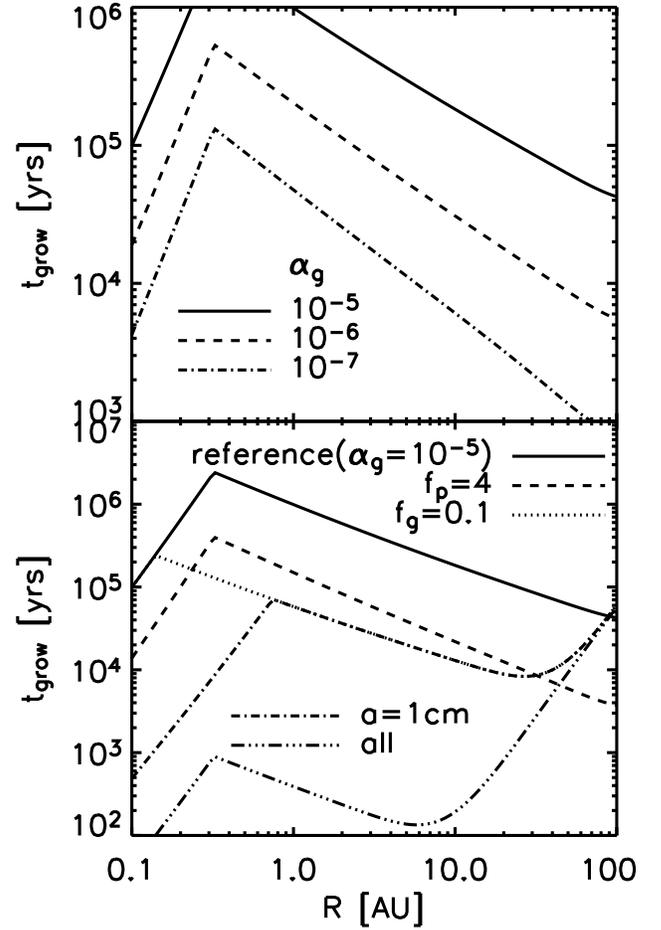} 
\vspace{-.2cm} 
   \caption{Growth times vs.\ radius for fixed $\ag$. \emph{Top:} our reference model ($f_g=f_p=\amm=1$) for $\ag = 10^{-5}$ (solid lines), $10^{-6}$ (dashed), and $10^{-7}$ (dot-dashed).   Sharp maxima denote a transition between drag laws.  \emph{Bottom:} with $\ag = 10^{-5}$, the disk model is varied by:  adding solids (dashed line), depleting gas (dotted line),  increasing particle size to 1 cm (dot-dashed line), and all of the above changes (triple-dot-dashed line).  The $f_g = 0.1$ model overlaps with the $a = 1$ cm curve for $\varpi \gtrsim 0.8$ and the reference model for $\varpi \lesssim 0.2$, because $\tau_s$ is identical in these regimes.}
   \label{fig:tgrow}
\end{figure} 

\begin{figure}[tb]
 \hspace{-.3in} 
 \includegraphics[width=3.8in]{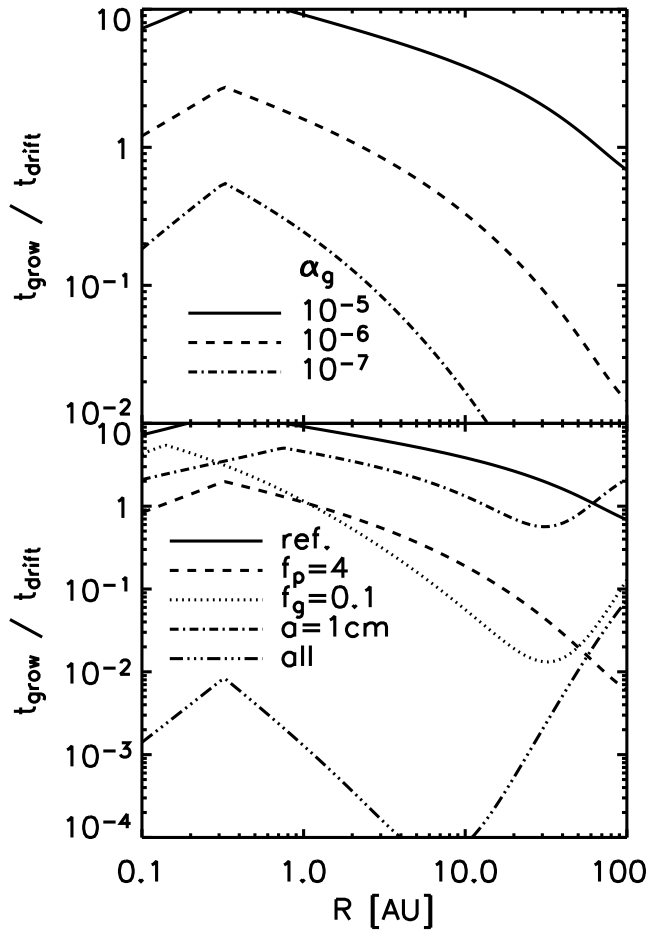} 
   \caption{Growth times vs.\ drift times for the same models as Figure \ref{fig:tgrow}.}
   \label{fig:gvd}
\end{figure} 

\section{Growth at Fixed $\ag$}\label{sec:fixalpha}
Figure \ref{fig:tgrow} (\emph{top}) plots the growth times in our reference model for several values of $\ag$.  The general results that growth is faster for weak turbulence and for $\tau_s$ near unity are confirmed.   The sharp peak at $\varpi  = R/\mathrm{AU} \approx 0.3$ occurs at the transition between Epstein's and Stokes' laws, where $\tau_s$ is minimized.

The detailed radial dependence is explained by the formulae in Table \ref{tab:limits} for $\tau_s \ll 1$:
\begin{equation}
{\tgr \over {\rm yr}} =
\begin{cases}
{\ag \Qm^2 \tau_s^{-1} \varpi^{3/2}  /(2\pi)}  & \text{if $\kfgm h \ll 1$} \\
\sqrt{\ag} \Qm \tau_s^{-3/2} \varpi^{3/2}  /(4\pi)  & \text{if $\kfgm h \gg 1$} \\
\end{cases}\, .
\end{equation}
Close to the star, in the Stokes regime, growth times decrease steeply, roughly as $\tgr \propto R^3$.  This is a combined effect of 
faster orbital times and shorter $\lmfp$, which weakens viscous drag coupling.  Moving radially outward from the drag transition, where Epstein drag applies, growth times also decrease. 
Slower orbital times are compensated by the decrease in $\tau_s$ at lower gas densities.  The radial decrease in $\Qm$, a result of our choice of power laws, also favors growth at large radii.    When $\tau_s > 1$ at sufficiently large radii,  $\varpi > 180(f_g/\amm)^{2/3}$,  looser coupling gives longer growth times.  Even if the disk is not truncated, rapid GIS will not occur at very large stellocentric distances.

In Figure \ref{fig:tgrow} (\emph{bottom}), $\alpha_s = 10^{-5}$ is fixed to show the effects of other model parameters on $\tgr$.  Enhancing the particle surface density via $f_p$ (dashed curves) gives stronger self-gravity and faster growth.
  Increasing particle sizes, $a$ (dot-dashed curves), or decreasing the gas surface density via $f_g$ (dotted curves) gives larger $\tau_s$.\footnote{Except $\tau_s^{\rm St}$ is independent of gas density.}  If $\tau_s < 1$  these changes shorten growth times, but once $\tau_s > 1$, the same changes give slower growth.  This explains the increase in $\tgr$ beyond $\sim 30$ AU in the $f_g = 0.1$ and $a = 1$ cm curves (and beyond $\sim 8$ AU when both changes are made).   When multiple effects favorable to GIS are included (gas depletion, larger particles, and/or particle enrichment) growth times are quite short in the inner disk, as seen in the curve labeled ``all." 

Figure \ref{fig:gvd} compares growth and drift times for the same models as Figure \ref{fig:tgrow}.  In many cases $\tgr > \tdr$, a violation of condition (\ref{driftcond}), though many of these modes have $\tgr > 10^5$ yrs as well.  The next section shows that drift rates do often set the most stringent constraint on $\ag$.  For now, note that the effects which give faster $\tgr$ also decrease $\tgr/\tdr$.  This is not immediately obvious since growth and drift times both decrease toward marginal coupling, but $\tgr$ does decrease faster.  Thus conditions which give faster growth also help growth outpace drift.

\begin{figure}[tb]
 \hspace{-.3in} 
 \includegraphics[width=3.8in]{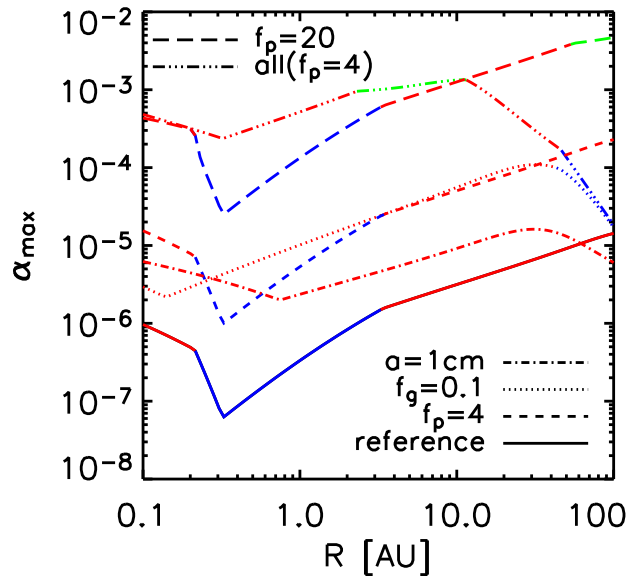} 
   \caption{Maximum levels of turbulent diffusion for which GI is relevant vs.\ disk radius for several disk models differentiated by linestyle.  Line colors indicate the limiting factor, red for radial drift, blue for disk lifetimes, and green for wavelengths comparable to disk radius.  See \S\ref{sec:amax}.}
   \label{fig:amax}
\end{figure} 

\begin{figure}[tbp]
 \hspace{-.3in}
  \includegraphics[width=3.8in]{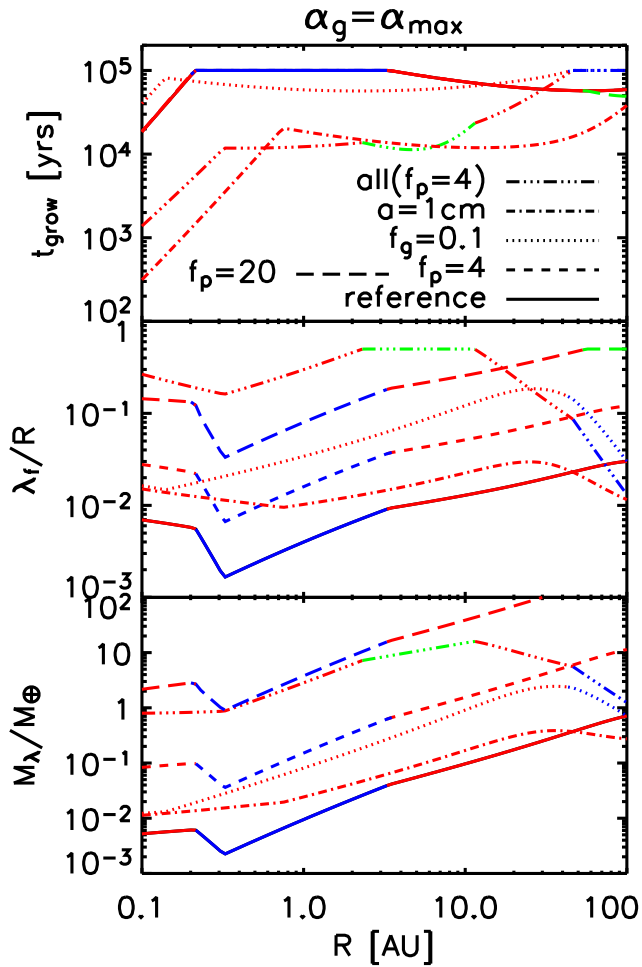} 

   \caption{Properties of unstable modes vs.\ radius for the strongest allowed turbulence.  Linestyles and colors have the same meaning as Figure \ref{fig:amax}.  Growth times (\emph{top}),  ratio of wavelength to disk radius (\emph{middle}), and mass of unstable modes in Earth masses (\emph{bottom}) are all maximum values that would be smaller for weaker turbulence. The reference, $f_p = 4$, and $f_p = 20$ models overlap on the top plot.} 
   
   \label{fig:tlm}
\end{figure} 

\section{Allowed Levels of Turbulence}\label{sec:alpha}
\subsection{Calculation of $\amax$}\label{sec:amax}
We now calculate $\amax$, the largest value of $\ag$ that satisfies all of conditions in equations (\ref{conditions}).  Figure \ref{fig:amax} plots $\amax$ against disk radius for the same models considered in Figure \ref{fig:tgrow}, plus a high enhancement, $f_p = 20$, case.\footnote{The $f_p = 20$ case is not well motivated in the outer disk where particle pile-ups have less of an influence.  Also the large total mass of solids would require efficient ejection from our solar system.}  At 1 AU, $10^{-6} \lesssim \amax \lesssim 10^{-3}$, showing that allowed levels of turbulence can be strong, but are model dependent.  Drift is often the limiting factor that determines $\amax$, as indicated by the line colors.   Since particle pile-ups naturally enhance the surface density of solids on $\tdr$ (YS02, YC04), they could act in concert with GIS.  A non-local model is required to explore this possibility in detail.  Drift also introduces shorter wavelength instabilities  which might concentrate particles faster than $\tdr$ \citep{gp00, yg05}, but their effect is not seen in the current single fluid, 2D analysis.

The disk lifetime (condition \ref{diskcond}) can set $\ag$ in two cases.  The first is near the Epstein-Stokes transition where stopping times are particularly short.  The second is in the outer regions ($R > 40$ AU) of gas depleted disks, where $\tau_s > 1$.  
The restriction that waves are shorter than the disk (condition \ref{lengthcond}) is relevant if conditions are very favorable to GIS, and $\amax$ is large.  For instance, the model ``all"  has wavelengths comparable to $R$ between $\sim 2$---10 AU.   

\subsection{Properties of ``Slowest Growing Modes"}\label{sec:slowmodes}
Figure \ref{fig:tlm} plots properties of unstable modes vs.\ $R$ for $\ag = \amax$.  This 
gives upper limits on growth times, wavelengths, and masses of unstable annuli, all of which would be smaller for weaker turbulence.  The top plot shows that growth times can be as long as $\sim 10^5$ yrs for $a = 1$ mm, but must be $\lesssim 10^4$ yrs for $a = 1$ cm.\footnote{Note that $\tdr$ (which equals $\tgr$ for the red [\emph{black in print version}] curves) is not decreasing $\propto R^{-1/2}$ as one expects for an individual particle.  This is due to greater particle settling and smaller $\mu_g$ at large $R$.}

Collision times could be shorter, but this does not, by itself, qualify as a weakness of the GI hypothesis when compared to collisional agglomeration.  First, we emphasize that we are quoting the longest allowed growth times.  Second, the collision time for large bodies is quite slow, especially in the outer disk:
\begin{equation}
t_\mathrm{coll} = {4 \rho_s a \over 3 \Sigma \Omega}{c \over \Omega h} \sim 10^4 {a \over \mathrm{km}} \varpi^3~\mathrm{yrs},
\end{equation}
where the numerical estimate uses $c \approx \Omega h$, appropriate for loose coupling (\S I.3.3).  Collision times will be shorter if a population of small bodies is maintained \citep{gls04}, but this requires fragmentation.  This brings us to the important point that collisional \emph{growth} times involve a balance between fragmentation and agglomeration.  With fragmentation included, it is not clear that the collisional growth time to km sizes is finite, let alone faster than $\tdr$ (very rapid in the 10 cm --- 1 m range) or disk lifetimes.  The collisional hypothesis is discussed further in \S\ref{sec:concl}.

The middle plot shows that wavelengths are not negligible compared to $R$.  This raises the possibility of  high resolution mm-wave observations, perhaps by ALMA.  Prospects would be particularly good for gas depleted or particle enhanced disks around $\sim 30$ AU.  For non-dissipative GI of solids, $\lambda_G/R \approx 10^{-5} f_p \varpi^{1/2}$, and there was little hope of detection. 

Upper limits on the mass of unstable modes are plotted in Figure \ref{fig:tlm} (\emph{bottom}).  These masses include a complete annulus, $M_\lambda = 2 \pi R \lfgm \Sigma$, because with $\tgr \gg \torb$ initial growth must be axisymmetric.  The reader should not be too concerned that these masses dwarf the ``canonical" values for kilometer-size planetesimals, $M_\lambda \sim \Sigma \lambda_G^2$, obtained from non-dissipative stability criteria \citep{gw73}.  

Non-linear effects should give subsequent fragmentation of the initial annulus.  As noted by \citet{war76}, when ``the local surface density increases, material becomes more susceptible to shorter-scale local dynamical instabilities."  Indeed the ring should fragment azimuthally, and perhaps radially, when $\tgr \lesssim \torb$.  Fragmentation will be enhanced by mass loading, the feedback of particle inertia on the gas turbulence (see \S\ref{alphacomp}), and the increasing rate of inelastic collisions.   As in star formation theory, the loss of support should lead to hierarchical fragmentation.    
Without a detailed understanding of these processes, we cannot estimate the planetesimal masses produced by GIS.

Nevertheless, by locking large amounts of solids in a mode for many orbital times, GIS would help preserve the chemical gradients in the solar nebula.  If drift compresses relative separations, as one would expect from particle pile-ups, then gradients could be amplified as well.  Implications for the asteroid belt are discussed in \S\ref{sec:concl}.

\subsection{Can Turbulence Prevent Gravitational Instability?}\label{sec:alphameaning}
We now compare our values of $\amax$, which range from $10^{-7}$ to $10^{-3}$ depending on disk model and radial position, to the expected levels of turbulence in protoplanetary disk midplanes.  
We first recall that previous analyses \citep[e.g.\ ][]{stu95} using the $Q_R \lesssim 1$ stability criteria placed stronger constraints,
\begin{equation}\label{alphaQR}
\ag \lesssim 5 \times 10^{-12} \varpi^{2} \amm{f_p^2 }/ f_g\, ,
\end{equation}
for tight coupling and Epstein's law.  Thus GIS is relevant for turbulence stronger by at least 5 orders of magnitude when dissipation is included.

\subsubsection{Turbulence Driven by Vertical Shear}
GIS involves particles settling to the disk midplane, and turbulence  arising from this stratification is well studied.  The source of instability is vertical shear in the azimuthal velocity.  Particle-free gas has an orbital speed that is slower than Keplerian by $\eta v_K$ due to radial pressure support.  As the gas is loaded with (tightly coupled) solids its orbital speed increases toward the Keplerian value.  Thus a vertical density gradient drives a velocity shear which can trigger turbulence as with the Kelvin-Helmholtz instability.  Semi-analytic analyses of these shear layers derive a thickness $h \approx \eta R/2$ in the tight coupling limit (\citealp{sek98}, YS02).  For looser coupling, one expects stronger turbulence and thinner particle layers, in agreement with \citet{cdc93}.

YS02 showed that only a critical surface density, $\Sigma_\mathrm{crit} \approx 2 \rho_g h$, of small solids can be stirred.  Any excess solids will form an overdense midplane layer, which would be subject to rapid GIS.   Since the current work (and Paper I) does not include mass loading effects, the saturation effect does not occur.  Thus our growth rates are probably too conservative, particularly for solid to gas ratios enhanced relative to cosmic abundances. 

In light of the current work, we ask whether stirred solids might also participate in GIS.  With $h \approx \eta R/2$, the parameter $Q_R \approx 250/f_p$.  The constant $Q_R$ curves of Figure \ref{fig:groworb} show that even with $f_p = 1$ marginally interesting growth is possible if $\tau_s > 10^{-2}$.  With some particle enhancement, and a lower $Q_R$, growth would be faster and relevant for more tightly coupled solids.  Thus GIS may take hold even if $\Sigma < \Sigma_\mathrm{crit}$.  Ongoing efforts to understand vertical shear instabilities \citep{gl04, go05} are improving understanding of this stirring mechanism and its effects on GIS.

\subsubsection{Accretion Related Turbulence}
More generally, protostellar disks are thought to be turbulent because mass accretion requires an angular momentum transport mechanism that is thought to be turbulent in nature.  The observed accretion rate of classical T-Tauri stars (CTTS) at 1 Myr is $\dot{M} \sim 10^{-8}~{\rm M}_\odot/{\rm yr}$ with order of magnitude scatter and a trend for $\dot{M}$ to decrease with age \citep{har98}.  These accretion rates are reproduced by viscous accretion disk models with $10^{-3} < \alpha_L < 10^{-1}$ \citep{d'a99}, where the diffusivity of angular momentum, $\alpha_L$, may differ from $\ag$ (see \S\ref{alphacomp}).   The relation between $\dot{M}$ (diagnosed observationally with UV veiling and emission line profiles) and $\alpha_L$ (from theoretical modelling) is not unique.  Thus the decrease in $\dot{M}$ with age could be due to weaker turbulence, lower disk masses, or a combination. 

The magneto-rotational instability is a leading candidate for the driver of angular momentum transport mechanism and disk accretion.  (Torques from density waves in massive gas disks are another possibility.)  Ideal MHD simulations indicate that the MRI generates $5\times 10^{-3} < \alpha_L < 0.5$ \citep{sto00}.  Much lower values are possible, including complete quenching, when resistive effects are important, i.e.\ the ionization fraction is too low.  In the ``layered accretion" scenario accretion flows are confined to ionized surface layers, leaving the resistive midplane relatively quiescent \citep{gam96}.

In summary there is observational and theoretical support for characteristic values of $10^{-1} \lesssim \alpha_L \lesssim 10^{-3}$ for angular momentum diffusion, but lower values are allowed.   We now consider the relation between $\alpha_L$ and diffusion coefficients relevant to GIS.

\subsubsection{Relation Between Diffusivities}\label{alphacomp}
For GIS we are mainly interested in the diffusivity of solids, $\alpha_s$.  Since particles respond to turbulent fluctuations of gas, we expressed our results in terms of $\ag$, using a Schmidt number ${\rm Sc} = \ag/\alpha_s$.  The relation between $\ag$ and $\alpha_L$, needed for comparison to stellar accretion rates, is characterized by a Prandtl number, ${\rm Pr} = \alpha_L/\ag$.  Additional relations are possible for anisotropic diffusion, to compare radial and vertical transport.

\citet{csp05} compared the radial diffusion of angular momentum and a passive contaminant (gas) in MHD turbulence and found ${\rm Pr} \sim 10$.  This result is readily explained by Maxwell stresses, which transport angular momentum without generating large velocity fluctuations.  This finding favors GIS.  However \citet{jk05} included vertical transport and particle coupling to measure $\alpha_L/\alpha_s = {\rm Pr}~{\rm Sc} \lesssim 1$.  These results appear to be a contradiction if one expects ${\rm Sc} > 1$, i.e.\ particles to be less diffusive than gas.

We used the \citet{cdc93} result for $\Sc$, equation (I.26), which is greater than unity due to the imperfect coupling of particles to eddies.   If particles preferentially concentrate in quiescent nodes of the turbulent flow \citep{cuz01}, then diffusion of solids would be less efficient, and  $\Sc$ would be even larger.  

Mass loading, the feedback of particle inertia on turbulence, should also decrease stirring efficiency.  Mass loading should become significant for $\rho \gtrsim \rho_g$ (YS02), or $Q_R \lesssim 140 \varpi^{-1/4}/f_g$.  While GIS is already significant at these densities without mass loading (see Fig. \ref{fig:groworb}), it is likely more rapid. Mass loading will certainly become significant as GIS develops, and the loss of turbulent support could lead to hierarchical fragmentation (see \S\ref{sec:slowmodes}).

To conclude, a variety of effects can decrease the efficiency of particle stirring in turbulent disks.  These include  mass loading, Maxwell stresses, layered accretion, and turbulent concentration. The role of these effects is only beginning to become clear.  Even with the pessimistic assumption that all the above effects are irrelevant and $\alpha_L = \alpha_g$,  disks must only be somewhat less active than characteristic CTTS values for GIS to be relevant.

\section{Discussion}\label{sec:concl}

We investigated the gravitational instability of solids (GIS) in protoplanetary disks with dissipation provided by gas drag including the stirring of  particles by turbulence.  GIS is faster for weak turbulence, high  abundances of solids, and particle stopping times comparable to orbital times.  We conclude that GIS is a viable mechanism to form planetesimals in disks that are not too strongly turbulent.  More importantly, as knowledge of physical conditions in disks improves through observation, theory, and simulation; the framework developed here can make firmer predictions about the role of GIS.  Possible refinements to our dynamical model were discussed at the end of Paper I.

An alternate hypothesis is planetesimals form by collisional agglomeration alone \citep{wc93}.  The physics of dust coagulation \citep{dt97} is sufficiently complex that predictions for growth beyond even $\mu$m sizes are difficult.  Existing theories do not include a realistic balance between constructive and erosive impacts over the relevant size range, from $\mu$m --- km.  Admittedly this a daunting task, and agglomeration cannot be ruled out simply because it computationally intractable.  However, as argued in YS02 and \citet{you04}, laboratory experiments and basic physical considerations do not give much reason for optimism.  Various mechanisms have been proposed to promote collisional growth.  For instance, gas flow through a porous body might return collisional fragments for repeat impacts at lower velocities \citep{wpk04, st05}.  Without a broader theory of coagulation in protoplanetary disks, the relevance of such sticking enhancement mechanisms is not clear.

Observational evidence for grain growth is stronger.  The most robust explanation for sub-mm spectral indices of protostellar disks is particle growth to sizes $\gtrsim 3$ mm \citep{cal02, c04,D05}.  This growth could be from coagulation of dry and/or icy grains, but might also involve partial melting, as in the formation of refractory meteoritic inclusions like chondrules \citep{shu01,dc02}.  Whatever the mechanism, growth to $\sim$ mm sizes aids settling to the midplane and abets GIS. 

We found that GIS develops as rings with radial wavelengths much longer than for traditional GI.  The large initial solid masses of unstable annuli does not necessarily imply large planetesimals would be formed.  Because subsequent fragmentation is not yet understood, we do not predict final planetesimal masses.  The potential combination of long growth times and AU-scale wavelengths increases the prospects for observation.  ALMA might image giant planets in the gaps they create \citep{wd05}.  Detecting a ring of solids would be more difficult, since sharp edges are not likely with a distribution of particle sizes.   

The implications of GIS for the asteroid belt should be considered.  Known meteorite falls from the asteroid belt (numbering $> 22,000$) are associated with a relatively small number, 100---150, of distinct ``parent bodies."  Some of these correspond to well-known asteroid classes.  It has been noted that the term ``parent body" could be misleading if many asteroids form from nearly identical material (see \citealp{bur02} for a review of meteoritic parent bodies).  The chemical homogeneity within a given group of asteroids might be due to  $\gtrsim 10^4$ years of mixing in a slowly collapsing ring prior to non-linear fragmentation into many planetesimals.

The strong chemical zonation between classes in the asteroid belt \citep[p.280]{tay01} might reflect the combined action of GIS, which segregates distinct annuli, and drift pile-ups (YS02, YC04).  By bringing neighboring rings closer together, pile-ups could steepen pre-existing compositional gradients.  These possibilities are rather speculative, but serve as a reminder that, despite 4.5 Gyr of collisional and dynamical evolution, the arrested development captured in asteroid belt provides invaluable clues for and tests of planetesimal formation theories.

Further circumstantial support connecting radial drift (but not necessarily GIS) to planetesimal formation is found in Jupiter's atmospheric abundances.  The uniform enrichment of noble gases and nitrogen points to a low temperature, $\sim 30$ K, origin for Jupiter's planetesimals \citep{owen99}.

The growth rates of GIS vary significantly with disk radius.  Slowest growth occurs near Mercury's orbit for mm-sized solids in a standard disk model.  The sharp decrease in growth times at closer distances to the star is encouraging for the \emph{in situ} formation of hot Neptunes and Jupiters, especially when there is a nearby stellar binary companion \citep{kon05}.  Growth times were also found to decrease toward the outer disk until decreasing gas density causes $\tau_s > 1$, and growth times increase.

This generally suggests that planetesimal formation at $\gtrsim 100$ AU (for mm-sized solids) might be difficult.  However it does not obviously predict that planetesimals should only form inside $\sim 50$ AU, as indicated by Kuiper Belt observations \citep{abm02}.  By contrast, YS02 and YC04 argued for a hard edge based on the pessimistic belief in threshold criteria for GI.
While it's a welcome change to have a theory that threatens to overproduce planetesimals, several scenarios could produce an edge inside $\sim 50$ AU.  Particles may migrate to the inner disk while turbulence is too strong for GIS.  Alternatively an external O star may rapidly photoevaporate the outer disk.

\acknowledgements 
This work benefitted greatly from helpful suggestions by Jeremy Goodman and Frank Shu.  I thank Bill Ward, Scott Tremaine, and Aristotle Socrates for stimulating discussions.  This material is based upon work supported by the National Aeronautics and Space Administration under Grant NAG5-11664 issued through the Office of Space Science.

\appendix

\section{A. Fast Eddies}\label{sec:fasteddy}
\begin{figure}[tb]
   \centering
 \includegraphics[width=4in]{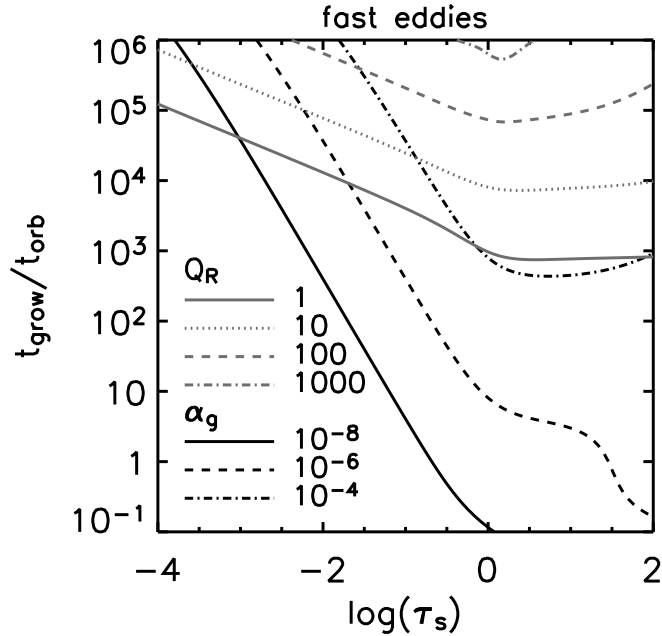}   
   \caption{Similar to Figure \ref{fig:groworb} except for faster eddies.  Linestyles have the same meaning as in Figure \ref{fig:groworb}.  Growth is slower for tight coupling but faster for loose coupling.  The drop in growth rates for $\tau_s > 10$ in the $\ag =10^{-6}$ curve occurs because $Q_T$ drops below unity (see Fig. I.3, \emph{bottom}). }
   \label{fig:fasteddy}
\end{figure}

In \S I.3.4.2, we considered the stirring of particles by fast eddies (FE) with short turnover times.      
Figure \ref{fig:fasteddy} plots growth times for this case, whose stability parameters were plotted in Figure I.3 (\emph{bottom}).   We compare to the growth rates found for orbital turnover times (OTT) in Figure \ref{fig:groworb}.
For growth with $\ag$ fixed (black curves), FE gives faster growth than OTT when $\tau_s > 1$.  This is easy to understand as both $Q_R$ and $Q_T$ are smaller for FE since loosely coupled particles respond poorly to rapid fluctuations.   

For $\tau_s < 1$, the growth is slower for FE than OTT.  Here coupling to the faster eddies is strong enough that velocity dispersion, and $Q_T$, are larger.  The higher densities (lower $Q_R$) for FE is less relevant because $Q_T^2/Q_R \gg 1$.  This corresponds to the thin disk, $\kfgm h \ll 1$, regime where growth is controlled by the velocity dispersion, not the density (see \S I.2.4).

However these growth time estimates for FE are probably too conservative.  We showed in \S I.C that collisional damping can be ignored when $c/(\Omega h) = Q_T/Q_R \ll 1$, which is the case for OTT (when $\tau_s < 1$).  But when $c/(\Omega h) \gg 1$, as is the case for FE (see \S I.3.4.2), we showed higher collision rates give significant dissipation.  This would give faster growth for FE.  Thus a more detailed analysis is required to determine if FE impedes or promotes growth (relative to OTT with constant $\ag$) for $\tau_s \ll 1$, but it clearly promotes growth for $\tau_s > 1$.
 
The growth times at fixed $Q_R$ (grey curves) are much longer for FE, simply because it takes much more vigorous turbulence (larger $\ag$ and thus $Q_T$) to stir solids to a given height with FE.  This demonstrates again that dissipative GI cannot be understood with a single stability parameter.  Note that the fluid approximation is questionable when $\tgr < \ts$, i.e.\ $\tgr/\torb < \tau_s/(2\pi)$, and a kinetic theory treatment would be more appropriate.  This technical comment applies to dynamical collapse in the lower right regions of both Figures \ref{fig:groworb} and \ref{fig:fasteddy}.  

\bibliography{refs}
\end{document}

%% file: tab1.tex
\begin{deluxetable}{llll}
\tablewidth{0pt}
\tablecaption{Symbols \label{tab:symb}}
\tablehead{\colhead{Symbol} & \colhead{Reference Eq.} & \colhead{Meaning}}

\startdata
$\Omega$, $\torb$ & $\torb = 2\pi/\Omega$ & orbital frequency, time\\
$\tgr$, $\gamma$ & eq.\ (\ref{eq:tgr}) & growth time, dimensionless rate \\
$\ts$, $\tau_s$ & eqs.\ (\ref{Ep},\ref{St}) & particle stopping time \\
$\tdr$&eq.\ (\ref{tdr}) & timescale for inward particle drift\\
$h$, $c$ & eqs. (I.29,I.30)& particle scale height, random speed\\
$Q_T$, $Q_R$ & eqs.\ (\ref{QT}, \ref{QR}) & stability parameters for solids\\
$\alpha_g$&eq. (\ref{eq:nu}) &turbulent diffusion parameter \\
$\Sigma$ & & particle surface density\\
$\rho$ & $\Sigma/h$ & particle space density\\
$a$, $\rho_s$ & & particle size, internal density\\
$f_p$, $f_g$ &  & particle, gas enhancement factors \\
$\kfgm$& $\kfgm = 2\pi/\lfgm$ & fastest growing  wavenumber\\
$\lfgm$& eq.\ (I.19) &fastest growing  wavelength\\
$\lambda_{G}$ & eq. (\ref{lg})& traditional GI wavelength\\
$M_\lambda$ & $2\pi \Sigma \lfgm R$ & mass of solids in unstable ring\\
$\eta$ & eq.\ (\ref{eta}) & pressure parameter\\
$R$, $\varpi$ & & cylindrical disk radius, in AU\\
$\rho_g$, $\Sigma_g$ & & gas space, surface density\\
$\mu_g$ & $\rho_g/(\rho_g + \rho)$ & midplane gas fraction\\
$\lmfp$& eq.\ (\ref{mfp}) & gas mean free path\\
$\Qm$& eq.\ (\ref{QsonicN}) & ``mixed" Toomre parameter\\
\enddata

\end{deluxetable} 

%% file: tab2.tex
\begin{deluxetable*}{llll}
\tablewidth{0pt}
\tablecaption{Limiting properties of unstable modes \tablenotemark{a} \label{tab:limits}}
\tablehead{\colhead{} & \multicolumn{3}{c}{Regime\tablenotemark{b}}\\
\cline{2-4}\\
\vspace{-.4cm}\\
\colhead{Quantity} & \colhead{$\tau_s \ll 1$, $\kfgm h \ll 1$} & \colhead{ $\tau_s \ll 1$, $\kfgm h \gg 1$} & \colhead{$\tau_s \gg 1$, $\kfgm h \ll 1$}
}

\startdata
$\Omega \tgr$ & $\ag \Qm^{2}/\tau_s$ & $\sqrt{\ag} \Qm\tau_s^{-3/2}/2$& $ \ag \Qm^{2}\tau_s $\\

$\lfgm /\lambda_G$ & $\ag \Qm^{2}$ & $\ag^{2/3} \Qm^{4/3}\tau_s^{-1/3}$& $\ag \Qm^{2}$\\

$\kfgm h$ & $(\ag \tau_s)^{-1/2}\Qm^{-1}$ & $(\ag \tau_s)^{-1/6}\Qm^{-1/3}$& $(\ag \tau_s)^{-1/2}\Qm^{-1}$\\
\enddata
\tablenotetext{a}{For orbital turnover times, $\tau_0 = 1$.}
\tablenotetext{b}{The $\tau_s \gg 1$,  $\kfgm h \ll 1$ case is not particularly relevant and was omitted.}
\end{deluxetable*}